\documentclass[a4paper]{mem}
\usepackage{natbib}
\usepackage{graphicx}
\begin{document}
   \title{A Multiband Approach to AGN:\\ Radioscopy \& Radio Astronomy 
}

   \author{M. Kadler \inst{1}, 
   E. Ros \inst{1}, 
           J. Kerp \inst{2},
   A.L. Roy \inst{1},
   A.P. Marscher \inst{3}
          \and
          J.A. Zensus \inst{1}\fnmsep
}

   \offprints{M. Kadler}
\mail{Auf dem H\"ugel 69, 53121 Bonn, Germany}


   \institute{Max-Planck-Institut f\"ur Radioastronomie,
Auf dem H\"ugel 69, 53121 Bonn, Germany \email{mkadler@mpifr-bonn.mpg.de}\\
              \and  Radioastronomisches Institut der Universit\"at Bonn, Auf dem H\"ugel 71, 53121 Bonn, Germany\\
              \and  Institute for Astrophysical Research, Boston University, 725 Commonwealth Avenue, Boston, MA 02215-1401, USA
             }

   \abstract{
Only in the radio-loud population of active galactic nuclei (AGN) does 
the production, collimation, 
and acceleration of powerful relativistic jets take place. 
We introduce here a concept of
combined VLBI- and X-ray spectroscopic observations of
sources with relativistic, broad iron lines. This approach has
enormous potential to yield deep insights into the
accretion/jet-production process in AGN.
Better knowledge of
the milliarcsecond-resolution radio structure of the
nuclear radio cores in so-called ``radio-quiet'' broad-iron-line
Seyfert galaxies is essential for future
combined radio/X-ray studies
of the different modes of radio-jet production in
accreting black hole systems.
   }
   \authorrunning{M. Kadler et al.}
   \titlerunning{Radioscopy \& Radio Astronomy}
   \maketitle
%

\section{Introduction}
The radioscopic\footnote{radioscopy: examination of the inner structure of opaque objects using X~rays or other penetrating radiation (WordNet 2.0, {\tt http://www.cogsci.princeton.edu/$\sim$wn})} 
study of relativistically broadened iron lines in the X-ray
regime 
allows the very closest neighbourhood of black holes
to be explored:
their accretion disks. These are thought to provide the ``fuel'' for the
jet production in radio-loud active galactic nuclei (AGN).
It is commonly accepted that broad iron lines
arise from
fluorescent K$\alpha$ emission
when the accretion disks of AGNs are irradiated by hard
X-rays from their coronae. 
Due to the strong relativistic effects in the inner disk, 
the iron-line profile is broadened and 
skewed (e.g., Fabian et al. 2000). 
If measured precisely enough, it reveals properties of the
accretion disk, in particular the 
orientation, extent and emissivity gradient radially outward from the 
black hole. Accretion events that might trigger enhanced jet-production
activity are expected to cause changes
in the line profile. Thus, combined jet and 
broad-iron-line monitoring can in principle
disclose the physical processes in accretion disks that lead to jet
production. 

While such broad iron lines from radio-quiet AGNs have been extensively studied
over the last decade, no significantly broadened lines have been detected in the
X-ray spectra of radio-loud AGN (Sambruna et al. 1999, Gambill et al. 2003),
which has been explained by either dominant beamed jet components, very high
ionization of the disks, or optically-thin,
radiatively-inefficient accretion flows.
Contrary to this expectation,
the first highly relativistic, broad iron line in a
radio-loud AGN was detected in NGC\,1052 (Kadler et al. 2004).
Not only is there a broad line, but
a change in the line profile coincides with an epoch of
jet-plasma ejection.

At present NGC\,1052 is the only strong radio source known
for which 
straightforward observational input can be gained from combined
high-resolution Very-Long-Baseline Interferometric (VLBI) 
observations and X-ray iron-line monitoring.
Thus, it is worthwhile to review the known radio-core properties of
the so-called ``radio-quiet'' broad-iron-line Seyfert galaxies.
We show here that these sources are not so quiet
after all at radio wavelengths but comprise a sample suitable for coordinated
X-ray and VLBI monitoring.

   \begin{figure*}
   \centering
   \resizebox{\hsize}{!}{\rotatebox[]{0}{\includegraphics{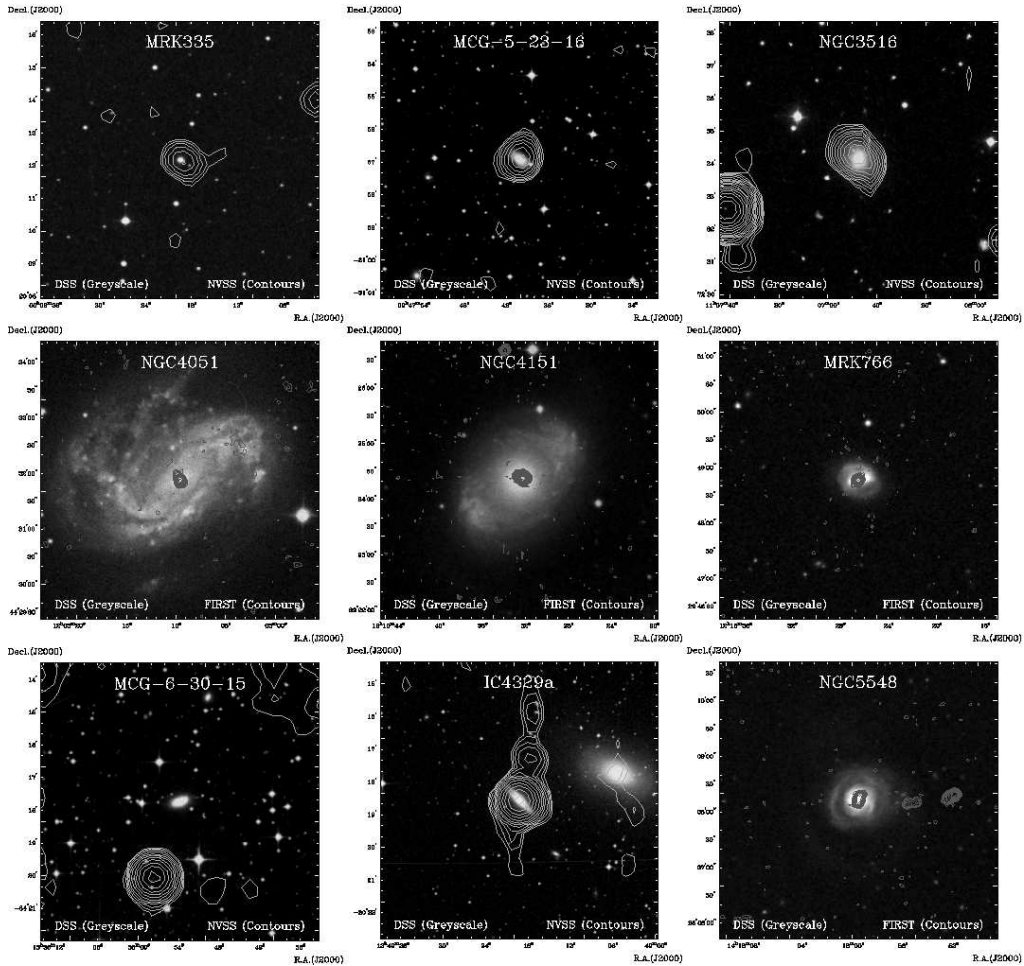}}}
   \caption{
Compact radio cores in ``radio-quiet'' Seyfert galaxies. 
Images display the optical 
host galaxies (DSS2 data), while contours show
the radio structure at 1.4\,GHz (FIRST 
and NVSS data, respectively). 
Integrated flux densities at $\lambda$6\,cm are given in Table~1.
              \label{fig:radio_cores}
}
    \end{figure*}

\section {The Compact Radio Cores in Broad-Iron-Line Seyfert Galaxies}
After the first detection of an extragalactic broad iron line 
associated with the Seyfert galaxy MCG\,$-$6$-$30$-$15
(Tanaka et al. 1995) 
there followed a
boom of follow-up {\it ASCA} detections in
other Seyfert galaxies and relativistic iron lines seemed to become a common
ingredient of Seyfert X-ray spectra.
However, in several cases 
{\it XMM-Newton}
could not verify the broad, relativistic red wings.
Firm {\it XMM-Newton} detections of broad iron lines have
been reported only for MCG\,$-$6$-$30$-$15, MCG\,$-$5$-$23$-$16, NGC\,3516,
Mrk\,335, and Mrk\,766 (see Table~1 for references).
It is not clear at present, whether insufficient
data quality from {\it ASCA} and over-simplified spectral models have led
to the discrepancy between the apparent presence of broad lines in
IC\,4329a, NGC\,4051, NGC\,4151, and NGC\,5548 or if intrinsic line
variability in these sources might have played a crucial role.
In Fig.~\ref{fig:radio_cores} we show the arcsecond-scale radio 
structure\footnote{\scriptsize Where available, we used radio images from the FIRST 
survey ({\tt http://sundog.stsci.edu}), otherwise we used images from the NVSS
{\tt (http://www.cv.nrao.edu/nvss})} 
of those Seyfert galaxies superimposed on optical images
from the 
Digitized Sky Survey\footnote{\scriptsize The Digitized Sky Survey was produced at the Space Telescope Science Institute under U.S. Government grant NAG W-2166. The images of these surveys are based on photographic data obtained using the Oschin Schmidt Telescope on Palomar Mountain and the UK Schmidt Telescope. The plates were processed into the present compressed digital form with the permission of these institutions.} 
compiled by using {\it Sky View}\footnote{\scriptsize \tt http://skyview.gsfc.nasa.gov}. 
All sources in this sample harbour compact radio cores at their 
centres. Only in the case of MCG\,$-$6$-$30$-$15 
does the flux density lie below
the detection limit of the NVSS. Several authors,
however, report the detection of a weak
but possibly highly variable flat-spectrum 
radio core in MCG\,$-$6$-$30$-$15, e.g., 
Ulvestad \& Wilson (1984) report 1.7\,mJy 
at $\lambda 20$\,cm whilst 
Nagar et al. (1999) measure 4.0\,mJy. 

\begin{table*}[th]
\caption{The sample}
\label{tab:sample}
\[
\centering
\resizebox{\textwidth}{!}{%
\begin{tabular}{@{}lcccp{6.5cm}@{}}
\hline
\hline
\noalign{\smallskip}
Source &   $S^\mathrm{VLA}_\mathrm{6cm}$ &  $F_{\rm (2-10)\,keV}$ & Broad & Notes \\
       &  {\scriptsize [mJy]}            & {\scriptsize [erg\,cm$^{-2}$\,s$^{-1}/10^{11}$]}  & Iron Line & \\
\noalign{\smallskip}
\hline
\noalign{\smallskip}
Mrk\,335            & 4.3$^\mathrm{a}$ & 1.7$^\mathrm{b}$& Yes$^\mathrm{c}$ & Seyfert 1.  Steep X-ray continuum with photon index of $\sim$2.3$^{(\mathrm{c})}$. Line profile might indicate a highly ionized, innermost region of the accretion disk around a rotating black hole.\\
MCG\,$-$5$-$23$-$16$^\mathrm{d}$ & 6.0$^\mathrm{a}$ & 7.5$^\mathrm{e}$ &  Yes$^\mathrm{e}$ & Seyfert 1.9.  Broad iron line emission, variable on timescales of months$^\mathrm{e}$. Time stable narrow fine-scale features. VLBI-unobserved.\\
NGC\,3516           & 4.3$^\mathrm{a}$ & 5.0$^\mathrm{b}$ & Yes$^\mathrm{f}$ & Narrow fine scale features indicate injection of material with a speed of 0.25\,$c$$^\mathrm{f}$.   \\
NGC\,4051           & 6.0$^\mathrm{a}$ & 2.0$^\mathrm{b}$ & Disputed$^\mathrm{g,h}$ & X-ray and radio variability$^\mathrm{i}$.  Correlation between the (accretion dynamics dominated) X-ray flux and the (jet dominated) radio flux. \\
NGC\,4151           & 125.0$^\mathrm{a}$&0.2$^\mathrm{b}$ & Disputed$^\mathrm{j}$ & \textsl{XMM-Newton} spectrum does not require a relativistically broadened iron line.  Narrow and variable Fe line, with amplitude of $\sim$25\%.  Compact jet$^\mathrm{k,l}$ \\
Mrk\,766            & 15.8$^\mathrm{m}$& 1.9$^\mathrm{b}$ & Yes$^\mathrm{n}$ & Broadened iron emission lines at 6.4\,keV and 6.7\,keV. Unconfirmed highly relativistic ``red wing"$^\mathrm{n}$.  Blueshifted Fe absorption edge at 8.7\,keV, suggesting ejected material at speeds of 15,000\,km\,s$^{-1}$ ($\beta\sim 0.05$). Only barely resolved with VLBI$^\mathrm{o}$\\
MCG\,$-$6$-$30$-$15 & 1.0$^\mathrm{a}$ & 3.8$^\mathrm{b}$ & Yes$^\mathrm{p}$ & The archetypical broad-iron-line galaxy. Broad and variable profile, fine structure$^\mathrm{q}$.  Compact VLA structure. VLBI-unobserved.\\
IC\,4329a           &  31.5$^\mathrm{m}$ &        8.3$^\mathrm{b}$ & Disputed$^\mathrm{r}$ & Disk possibly truncated shortly before reaching the innermost stable orbit around the central black hole$^\mathrm{s}$. VLBI-unobserved.\\
NGC\,5548           & 10.5$^\mathrm{m}$& 6.0$^\mathrm{t}$ & Disputed$^\mathrm{t}$ & Soft excess varies more strongly than the high-energy continuum$^\mathrm{u}$.   The spectrum shows reflection and fluorescence from neutral iron distant from the central source$^\mathrm{t}$.  Rapid radio variability$^\mathrm{v}$.  \\
\noalign{\smallskip}
\hline
\end{tabular}
}
\]
\begin{scriptsize}
\begin{list}{}{
\setlength{\leftmargin}{0pt}
\setlength{\rightmargin}{0pt}
}
\item[]
$^\mathrm{a}$ Ulvestad \& Wilson 1984;
$^\mathrm{b}$ HEASARC website: {\tt http://heasarc.gsfc.nasa.gov/};
$^\mathrm{c}$ Gondoin et al. 2002;
$^\mathrm{d}$ See Fabian et al. 2000 for a review;
$^\mathrm{e}$ Dewangan et al. 2003;
$^\mathrm{f}$ Turner et al. 2002;
$^\mathrm{g}$ Wang et al. 1999;
$^\mathrm{h}$ Pounds et al. 2004;
$^\mathrm{i}$ McHardy et al. (2004, \& priv. comm.); 
$^\mathrm{j}$ Schurch et al. 2003;
$^\mathrm{k}$ Ulvestad et al. 1998; 
$^\mathrm{l}$ Mundell et al. 2003;
$^\mathrm{m}$ Rush et al. 1996;
$^\mathrm{n}$ Pounds et al. 2003b;
$^\mathrm{o}$ Lal et al. 2004;
$^\mathrm{p}$ Fabian et al. 2002;
$^\mathrm{q}$ Wang et al. 2004;
$^\mathrm{r}$ Gondoin et al. 2001;
$^\mathrm{s}$ Done et al. 2000;
$^\mathrm{t}$ Pounds et al. 2003a;
$^\mathrm{u}$ Done et al. 1995;
$^\mathrm{v}$ Wrobel 2000
\end{list}
\end{scriptsize}
\end{table*}

\section{Multiband-Scrutiny of Broad-Iron-Line Galaxies}
Combined VLBI (radio astronomical) and X-ray spectral (radioscopic)
observations of AGN that exhibit both a strong broad iron line and a
compact radio jet can study 
the coupling of mass accretion and jet formation around supermassive black 
holes. 
Processes that trigger jet production are expected to 
leave clear marks in the time-dependent relativistic-iron-line profile, which
is sensitive to changes of the inner edge of the accretion
disk and so is possibly the most direct tracer of dynamical processes in
the accretion flow. More complicated models might invoke
changes in the structure of the magnetic field in and above the disk.
According to these models,
turbulent field corresponds to higher disk viscosity while predominantly poloidal field
facilitates flow of energy into the jet 
(e.g., Livio et al. 2003). 
Tagger et al. (2004) propose that there is a period of mainly poloidal field 
during which the accretion flow moves
toward smaller radii until it reaches the innermost orbit.
At this point reconnection of the field
lines can inject energy into the jet explosively 
(Eikenberry \& van Putten 2003).
Alternatively, instabilities in the
inner disk could cause the accretion flow to be irregular 
(Belloni 2001),
with ejection of extra
material into the jet occurring when a chunk of matter suddenly falls inward. 
Such scenarios 
are under active development to build
models that will predict the variation of the iron line profile
together with that of the X-ray continuum.  
We are beginning
combined VLBI and X-ray monitoring observations of these
broad-iron-line AGN to
yield
the data needed to test such models.

\begin{acknowledgements}
{\scriptsize
M.\,K. was supported for this research through a stipend from the International
Max Planck Research School (IMPRS) for Radio and Infrared Astronomy at the University of Bonn.
}
\end{acknowledgements}

\bibliographystyle{aa}

\end{document}